\documentclass[]{emulateapj}
\usepackage{apjfonts}
\usepackage{psfig}

\shorttitle{}
\shortauthors{}
\begin{document}
\title{The Highest Dynamical Frequency in the Inner Region of an Accretion Disk}
\author{M.\ Ali Alpar}
\affil{Faculty of Engineering and Natural Sciences,Sabanc{\i} University, Orhanl{\i}, Tuzla, Istanbul 34956, Turkey}
\email{alpar@sabanciuniv.edu}

\and 
\author{Dimitrios Psaltis\altaffilmark{1}}
\affil{Physics and Astronomy Departments, University of Arizona, 
1118 E.\ 4th St., Tucson, AZ 85719}\email{dpsaltis@physics.arizona.edu}
\altaffiltext{1}{also, Visiting Professor, Sabanc{\i} University}

\begin{abstract} In the inner regions of accretion disks around
compact objects, the orbital frequency of the gas deviates from the
local Keplerian value. For long-wavelength modes in this region, the
radial epicyclic frequency $\kappa$ is higher than the azimuthal
frequency $\Omega$. This has significant implications for
models of the twin kHz QPOs observed in
many neutron-star sources that traditionally identify the frequencies
of the two kHz QPOs with dynamical frequencies in the accretion disk.
The recognition that the highest frequency in the transition or
boundary region of the disk is actually the epicyclic frequency also
modifies significantly the constraints imposed by the observation of 
high-frequency QPOs on the mass and radius of the compact objects.
\end{abstract}
\keywords{accretion disks ---  waves}

\section{INTRODUCTION}

The X-ray brightness of an accreting compact object is often observed
to be modulated quasi-periodically at different characteristic
frequencies that are comparable to the dynamical timescale of the
central object (see, e.g., van der Klis 2005). The physical origin of
the various types of quasi-periodic oscillations (QPOs) in accreting
black holes and neutron stars are still a matter of debate (see, e.g.,
Psaltis 2004; van der Klis 2005). Indeed, different variability models
attribute the observed QPOs to brightness variations generated at
different dynamical frequencies at a particular radius in the
accretion disk (e.g., Alpar \& Shaham 1985; Miller, Lamb, \& Psaltis
1998; Stella, Vietri \& Morsink 1999; Abramowicz et al.\ 2003) or
at the frequencies of wave modes in the disk, which are also related to the
dynamical frequencies (e.g., Alpar et al.\ 1992; Alpar \&
Y{\i}lmaz 1997; Wagoner 1999; Kato 2001).

Despite their differences, most proposed models agree that the highest
frequency of a large amplitude QPO cannot be larger than the frequency
of a stable dynamical oscillation in the accretion disk. Around a
non-rotating black hole or a relatively compact neutron star of mass
$M$, the highest stable dynamical frequency is the azimuthal orbital
frequency at the radius of the innermost stable circular orbit
\begin{equation} 
f_{\rm ISCO}=\frac{c^3}{12\pi\sqrt{6}GM}\;.
\label{eq:isco} 
\end{equation}
Requiring this frequency to be larger than the highest observed QPO
frequency imposes an upper bound on the mass of the compact object
(e.g., Miller, Lamb, \& Psaltis 1998) or even provides evidence that
the compact object is rapidly spinning (e.g., Strohmayer 2001).

In a realistic accretion disk, the dynamical frequencies of
oscillations of fluid elements are approximately equal to the
dynamical frequencies of test particles (such as eq.~[\ref{eq:isco}])
only away from the boundaries. For example, in the accretion disk
around a neutron star with a dynamically important magnetic field, the
orbital frequency of a fluid element near the so-called Alfv\'en
radius (see, e.g., Ghosh \& Lamb 1991) deviates significantly from the
local Keplerian frequency because of magnetic stresses and viscous stresses
(Erkut \& Alpar 2005).  In accretion
flows without large scale magnetic fields, such as those around a
non-magnetic neutron star or a black hole, the Maxwell stresses due to
turbulent small-scale magnetic fields can also affect the dynamical
frequencies in the inner disk (see, e.g., Hawley \& Krolik 2001). Even
in the absence of any magnetic fields, radiation drag forces can alter
significantly the dynamical frequencies at the inner regions of
accretion disks (Miller \& Lamb 1996).

The magnetic and radiation forces alter the dynamical frequencies
predominantly near the inner regions of the Keplerian flows, where the
observed QPOs are expected to originate. In this {\em Letter}, we
argue that in realistic accretion disks, in which the azimuthal
orbital frequency has a maximum at some radius outside the surface or
horizon of a compact object, the radial epicyclic frequency at a
comparable radius is the highest dynamical frequency in the
system. Our result has significant implications for models of QPOs in
accreting compact objects and for the constraints imposed on the
masses and spins of the compact objects by the observation of
high-frequency QPOs.

\section{THE HIGHEST DYNAMICAL FREQUENCY}

In order to discuss the relative order of different dynamical
frequencies in an accretion disk independently of the details of the
additional forces that affect the motion of fluid elements, we use the
simple but transparent derivation of the equations of motion discussed
by Hill (1878). These equations are valid for the motion of test
particles in a frame rotating with an angular velocity $\Omega_0$, in
the presence of additional forces. This set of equations has been
used by Balbus \& Hawley 1992 to elucidate the physics behind the
magnetorotational instability in MHD accretion disks (see also Pessah
\& Psaltis 2005 for a discussion of a more general case of MHD
instabilities).

We start with particles in a stable circular orbit at a radius $r_0$ and
with an angular velocity $\Omega_0=\Omega(r_0)$, i.e., $r=r_0$, 
$\phi = \Omega_0 t $, where $\Omega(r)$ is not necessarily the Keplerian
angular velocity. Introducing small perturbations around the circular 
orbit, i.e., $r = r_0 + x$ and $\phi = \Omega_0 t + y/r_0$,
to first order in $x$ and $y$, we obtain the Hill equations
\begin{eqnarray} 
\ddot{x} - 2 \Omega_0 \dot{y} & = & - x \left[r \frac{d\Omega^2}{dr}\right]
_0+f_x \\ 
\ddot{y}+ 2 \Omega_0 \dot{x} & = & f_y\;,
\end{eqnarray} 
where $f_x$ and $f_y$ represent the $x-$ and $y-$ components of the sum
of the perturbations of all forces besides gravity that are acting on
the particles. For small displacements away from the stable equilibrium
orbit, $f_x$ and $f_y$ are linear in $x$ and $y$ with negative first
derivatives, and hence
\begin{eqnarray} 
\ddot{x} - 2 \Omega_0 \dot{y} & = & - x \left[r \frac{d\Omega^2}{dr}\right]_0
-\left\vert\frac{\partial f_x}{\partial x}\right\vert_0x
-\left\vert\frac{\partial f_x}{\partial y}\right\vert_0y \nonumber\\ 
\ddot{y}+ 2 \Omega_0 \dot{x} & = & 
  -\left\vert\frac{\partial f_y}{\partial x}\right\vert_0x
-\left\vert\frac{\partial f_y}{\partial y}\right\vert_0y\;.
\label{eq:hill}
\end{eqnarray}

Looking for solutions of the form $\sim {\rm e}^{-i\omega t}$
we obtain the dispersion relation
\begin{eqnarray}
& &\omega^4-\left(\kappa^2+\left\vert\frac{\partial f_x}{\partial x}\right\vert_0
+\left\vert\frac{\partial f_y}{\partial y}\right\vert_0\right)\omega^2
-2i\Omega_0\left(\left\vert\frac{\partial f_y}{\partial x}\right\vert_0
-\left\vert\frac{\partial f_x}{\partial y}\right\vert_0\right)\omega
\nonumber\\
& &+\left(\left[\frac{d\Omega^2}{d\ln r}\right]_0+
    \left\vert\frac{\partial f_x}{\partial x}\right\vert_0\right)
 \left\vert\frac{\partial f_y}{\partial y}\right\vert_0
 -\left\vert\frac{\partial f_y}{\partial x}\right\vert_0
 \left\vert\frac{\partial f_x}{\partial y}\right\vert_0=0\;,
\label{eq:disp}
\end{eqnarray}
where
\begin{equation}
\kappa^2\equiv 4\Omega_0^2+\left[\frac{d\Omega^2}{d\ln r}\right]_0\;.
\label{eq:kappa}
\end{equation}
With axial symmetry, $\partial f_x/\partial y=\partial f_y/\partial y=0$,
\begin{equation}
\omega^4-\left(\kappa^2+\left\vert\frac{\partial f_x}{\partial x}\right\vert_0\right)\omega^2
-2i\Omega_0\left\vert\frac{\partial f_y}{\partial x}\right\vert_0\omega=0\;.
\label{eq:sym}
\end{equation}
When non-gravitational forces are absent, the solutions are stable modes
with $\omega=0$ and $\omega=\kappa$. As can be verified easily,
$\omega = \kappa$ is the frequency associated with the mode of radial
oscillations.  As azimuthal perturbations simply shift the phase angle
$\phi$ among equivalent phases on the stable circular orbit
("spontaneous symmetry breaking"), there is no restoring force, and 
the solution corresponding to azimuthal oscillations is simply
$\omega=0$. Indeed, after an azimuthal
perturbation the particle would proceed from its shifted phase on the
same circular orbit, at the original orbital frequency
$\Omega_0$. This is an instance of the Goldstone theorem that zero
frequency modes are associated with spontaneous symmetry breaking.
The physical reason for the nonzero restoring force and nonzero
frequency $\omega = \kappa$ for radial perturbations is rotation.  The
long range correlations introduced by the presence of rotation,
i.e., the Coriolis force and the gradient of the centrifugal potential
provide the restoring force in the Hill equations~(\ref{eq:hill}). Modes
effected by long range forces always have nonzero frequencies, as
exemplified by the plasma frequency due to the Coulomb force, the
Jeans mass associated with gravitational instabilities, and the finite
mass of Higgs bosons.

The effect of short-range forces can be traced in the realistic limit
\begin{equation}
\left\vert\frac{\partial f_x}{\partial x}\right\vert_0,
\left\vert\frac{\partial f_x}{\partial y}\right\vert_0,
\left\vert\frac{\partial f_y}{\partial x}\right\vert_0,
\left\vert\frac{\partial f_y}{\partial y}\right\vert_0
\ll \kappa^2\;.
\label{eq:cond}
\end{equation}
The $\omega\simeq \kappa$ modes are now
quasi-stable with ${\cal R}(\omega)\simeq \kappa$ and ${\cal
I}(\omega) \sim {\cal O}(\vert\partial f_{x,y}/\partial
x\vert_0/\kappa)\ll {\cal R}(\omega)$.  There is still an exact
$\omega=0$ mode, due to the axial symmetry and a 
decaying mode with ${\cal R}(\omega)=0$ and ${\cal I}(\omega) \sim - {\cal
O}(\kappa)$. This decay corresponds to the effect of the
shear on the initial perturbations.

When the disk is treated as a continuous medium and its hydrodynamic
or magnetohydrodynamic wave modes are studied, the analogues of
restoring forces on perturbations of single particle orbits in the
Hill equations are spatial derivatives of the forces on a fluid
element. These restoring force terms contain the product
$(\vec{V}_A\cdot\vec{k})^2$ or $(c_s k)^2$ where $\vec{k}$ is the wave
vector, and $\vec{V}_A$ and $c_s$ are the Alfv\'en velocity and sound
speed respectively (see, e.g., Alpar et al.\ 1992; Balbus \& Hawley
1992; Alpar \& Y{\i}lmaz 1997; Kato 2001; Pessah \& Psaltis
2005). Conditions analogous to (\ref{eq:cond}) describe global modes
of long wavelength. These modes are expected to be the ones that
produce the largest amplitude of brightness modulation.  In the limit
of small wavenumber $k$, the frequencies of oscillations have values
close to $\kappa$ and $\Omega$. For modes with m azimuthal nodes,
frequencies comparable to $\kappa - m\Omega$ will appear; the dominant
global modes will be those with $m=0$ and $m=1$, which represents the
beat between the radial epicyclic frequency and the disk rotation.
This fundamental effect of rotation can be seen in all dispersion
relations of waves in rotating fluids and accretion disks, in models
of varying degrees of complexity (see, e.g., Chandrasekhar 1961;
Papaloizou \& Pringle 1984).  We emphasize here that not all
oscillations described by the Hill equations are stable. This is of
particular relevance to MHD disks that are subject to the
magnetorotational instability, especially at the limit of small
wavenumbers.

Where magnetic or radiation forces significantly
affect the orbital motion of gas elements, the azimuthal
frequency in the disk differs from the local Keplerian
value. As an example, for accretion onto a rotating and magnetic
neutron star, the local azimuthal
frequency $\Omega(r)$ is lower than the Keplerian frequency inside some
characteristic radius $r_2$ comparable to the Alfv\'en radius. It
has a maximum $\Omega_{\rm max}$ at $\bar{r}<r_2$ and then decreases to match 
the rotation frequency of the star $\Omega_*$ at a radius $r_1$, which
marks the disk-magnetosphere boundary. The simplest mathematical model 
employs a quadratic form for $\Omega(r)$,
\begin{equation}
\Omega(r) =
\left\{
\begin{array}{ll}
 \Omega_{\rm max} - (\Omega_{\rm max} - \Omega_*)(\frac{r - \bar{r}}{r_1 - \bar{r}})^2\;, & r\le r_2\\
 \left(GM/r^3\right)^{1/2}\;, & r>r_2
\end{array}
\right.\;,\label{eq:model}
\end{equation}
where $M$ is the mass of the star. Matching these two expressions, as well
as their derivatives, at $r_2$ relates the maximum frequency and its location
to the transition region parameters $r_1$, $r_2$; and the neutron-star rotation rate $\Omega_*$:
\begin{equation}
\Omega_{\rm max}=\frac{\Omega_*/\Omega_2-(9/16)(r_1/r_2-7/3)^2}
{\Omega_*/\Omega_2+(3/2)(r_1/r_2-5/3)}\Omega_2=\frac{7}{4}\left(1-\frac{3}{7}\frac{\bar{r}}{r_2}\right)\Omega_2\;.
\end{equation}
This model quite accurately represents the $\Omega(r)$ curves that
meet a boundary condition at the stellar magnetosphere and
asymptotically join the Keplerian curve within a transition region in
models of accreting neutron stars with dynamically important magnetic
fields Erkut \& Alpar 2005.  The model of the azimuthal frequency
profile described by equation~(\ref{eq:model}) is shown in Figure~1,
for $\Omega_2/\Omega_*=1.1$, $\Omega_{\rm max}/\Omega_*=1.2$,
and $r_1/r_2=0.84$, together with the radial epicyclic frequency
calculated according to equation~(\ref{eq:kappa}).

At radii $r<r_2$, where the azimuthal frequency is lower than the
Keplerian frequency, the radial epicyclic frequency is the larger of
the two dynamical frequencies. At the radius $\bar{r}$, where the
azimuthal frequency has a maximum, $\kappa=2\Omega_{\rm max}$ and the
ratio $\kappa/\Omega$ increases to even larger values in the region
$r_1<r<\bar{r}$, where the azimuthal frequency is decreasing with
decreasing radius.  For a wide range of radii inside $r_2$, the radial
epicyclic frequency is larger than the local Keplerian frequency. As a
result, if low-$k$ modes in this transition region are responsible for
the observed QPOs in accreting neutron stars and black holes, as
envisioned by most models, then an upper bound of the QPO frequencies
may be set not by the maximum azimuthal frequency but by the maximum
radial epicyclic frequency in the region.

\begin{figure}[t]
 \centerline{
\psfig{file=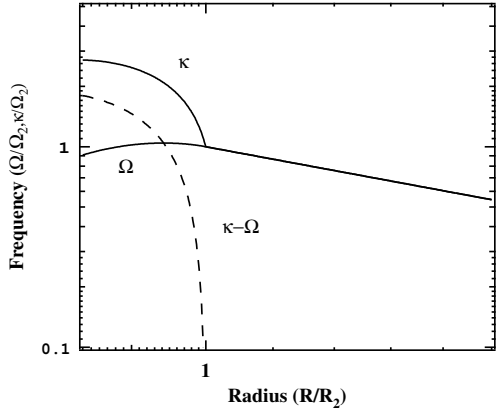,angle=0,width=7truecm}}
  \caption{The radial profile of the azimuthal ($\Omega$) and radial
epicyclic ($\kappa$) frequencies of the simple model discussed in the text,
for $\Omega_*/\Omega_2=1.1$ and $r_1/r_2=0.4$. The dashed line shows
the difference $\kappa-\Omega$. At radii $r>r_2$ the azimuthal and
radial epicyclic frequencies are equal to the Keplerian frequency.}
\end{figure}

\section{DISCUSSION}

In \S2, we argued that, in the inner regions of accretion disks around
compact objects, magnetic, radiation, pressure, and viscous forces may
become comparable to the gravitational force, alter the orbital
frequencies of fluid elements and lead to radial epicyclic frequencies in
excess of the orbital frequencies. Here, we discuss the implications of
our results for models of the kHz QPOs observed from many accreting
neutron-star sources (for a review see van der Klis 2005).  These are
pairs of QPOs with frequencies comparable to a kHz that vary on
timescales longer than any of the dynamical timescales in the vicinity
of the neutron stars. 

As the frequencies of the kHz QPOs vary, they follow a number of
intriguing patterns. In all sources for which the spin frequency of
the neutron star is known, either via observations of X-ray pulses or
burst oscillations, the difference frequency between the two kHz QPOs
was shown to be comparable to the neutron star spin frequency or to
half its value. This property gave rise to the beat-frequency models
of kHz QPOs (see Strohmayer et al.\ 1996; Miller et al.\ 1998;
Chakrabarty et al.\ 2003). In sources for which the frequencies of the
two kHz QPOs were measured with high accuracy, such as Sco~X-1, it was
shown that the two frequencies followed a quadratic correlation
(Psaltis et al.\ 1998; Psaltis, Belloni, \& van der Klis 1999).
Together with the correlation between the frequencies of the kHz QPOs
and other low-frequency QPOs observed simultaneously in the same
sources, this provided support for the relativistic precession models
(Stella \& Vietri 1999; Stella, Vietri, \& Morsink 1999; Psaltis \&
Norman 2000). Finally, the ratios of the frequencies of the two kHz
QPOs are roughly comparable to the ratio $2/3$ and this gave rise to
models that rely on resonances between modes (Abramowicz et al.\
2003). It is important to note here that, mathematically speaking, not
all three of the above patterns can characterize simultaneously the
frequency correlations of the kHz QPOs and, in fact, none of them
is valid to within the measurement errors of the QPO frequencies.
However, all three of them can be shown to be able to describe the
data roughly, typically to within 30\% for the first and last
alternative and to within 5\% for the quadratic correlation (see
Psaltis et al.\ 1998 for a detailed discussion of the statistical
significance of various correlations).

In all the models of the kHz QPOs mentioned above, one or both of the
observed QPO frequencies are interpreted as dynamical frequencies
(azimuthal, radial, or vertical) at different characteristic radii in
the accretion flows. For each of these models, these dynamical
frequencies have been calculated so far assuming that the only force
that affects the motion of fluid elements is gravity, even though
there is always an implicit assumption that some additional physical
mechanism picks the characteristic radius where these QPOs
originate. As we have shown, the relative ordering of the frequencies
for long-wavelength modes depends on the radial profile of the
azimuthal frequency. Therefore, the identification of observed QPOs
with particular dynamical frequencies may not be self consistent in
any of the above models.

For most realistic models of the azimuthal frequency, e.g., for the
quadratic model discussed in \S2 (eq.~[\ref{eq:model}]), the largest
of the dynamical frequencies is $\kappa$ and the smallest is $\Omega$,
in the inner part of the transition region where d$\Omega/dr > 0$ (see
also Fig.~1). Wavepackets in the disk can modulate the accretion flow
onto the compact object at the radial epicyclic frequency band $\sim
\kappa$, at the orbital frequency band $\sim \Omega$, or at the beat
$\kappa- \Omega$ of the radial and orbital motions.  It is clear
that $\kappa \geq 2 \Omega$ and therefore $\kappa - \Omega \geq
\Omega$ in the region d$\Omega/dr > 0$, where the order of the
frequencies is $\kappa > \kappa - \Omega \geq \Omega$.

How does one identify the observed upper and lower kHz QPO
frequencies, hereafter $\nu_2$ and $ \nu_1$, respectively, with two of
these three possible frequencies? There are two simple trends that all
kHz QPOs are observed to obey: (i) the two kHz QPO frequencies
always increase or decrease together, and (ii) their difference
$\Delta \nu \equiv \nu_2 - \nu_1$ decreases as both frequencies
increase.  The only pair of frequencies that satisfies these two
conditions are $\kappa$ and $\kappa - \Omega$. Indeed, if we associate
these two frequencies with the upper and lower kilohertz QPO
frequencies, i.e., $\kappa$ = 2$\pi \nu_2 $ and $\kappa - \Omega$ =
2$\pi \nu_1 $, they will satisfy the trend (ii) if
$d\Omega/d\kappa<0$, which is equivalent to 
\begin{equation}
\frac{d\log(\Omega d\Omega/dr)}{d\log r} < -5\;.  
\end{equation} 
This condition holds as long as the transition region size is about a
fifth of the disk inner radius or less. 

Perhaps the most interesting application of kilohertz QPO models is
the constraints they provide on the mass-radius relation of neutron
stars (Miller et al.\ 1998; see also discussion after Eq.~[1]).  The
interpretation of the highest QPO frequency as the epicyclic rather
than the Keplerian azimuthal frequency modifies the bounds obtained
from the upper kilohertz frequency. To estimate the effect of the
phenomena discussed here on the bounds in the mass-radius plane, we
use the simple model~(\ref{eq:model}).  In this model, the maximum
epicyclic frequency, to a first approximation, is equal to the radial
epicyclic frequency at radius $\bar{r}$ and hence \begin{equation} 2
\pi \nu_{2,{\rm max}} \simeq \kappa(\bar{r}) = 2 \Omega_{\rm max} \;.
\end{equation}

\begin{figure}[t] 
\centerline{
\psfig{file=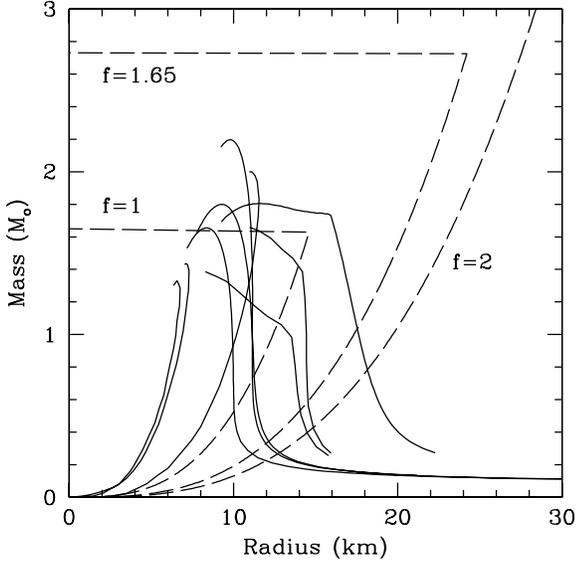,angle=0,width=8truecm}} 

\caption{Constraints on the mass-radius plane of neutron stars imposed
by the observation of the highest frequency QPO so far (1330~Hz, see
van Straaten et al.\ 2000) for different values of the parameter $f$ (see,
eq.~[\ref{eq:f}]). The solid lines show the predictions of
representative equations of state for neutron-star matter (see
Lattimer \& Prakash 2001).}
 \end{figure}

The constraints the kHz QPOs impose on the mass and the radius of a
neutron star are obtained from two requirements. First, the neutron
star radius $r_{\rm NS}$ must be less than a representative inner disk 
radius that is associated with upper kilohertz QPO frequency, which was 
interpreted as a Keplerian frequency in earlier applications
(see van der Klis 2005). Recognition of the epicyclic frequency as the highest
frequency modifies this constraint to
\begin{equation} 
r_{\rm NS} < \frac{(GM)^{1/3}} {(2 \pi \nu_2)^{2/3}} f^{2/3}\;,
\end{equation} 
where the factor 
\begin{equation} 
f \equiv \left[\frac{\kappa(\bar{r})}{\Omega_K(\bar{r})}\right]\simeq 
\frac{7}{2}\left(\frac{\bar{r}}{r_2}\right)^{3/2} 
\left(1-\frac{3}{7}\frac{\bar{r}}{r_2}\right) 
\label{eq:f}
\end{equation} 
depends on the width of the transition region.  The second constraint
simply states that the radius of the last stable orbit is less than
the disk radius associated with the upper kilohertz QPO,
$r_{\rm ISCO}= 6GM/c^2<\bar{r}$. This leads to
\begin{equation} 
M < \left(\frac{c^3}{6\sqrt{6}G}\right) \frac{f}{2 \pi \nu_2}\;,
\end{equation} 
if the upper kilohertz QPO is the radial epicyclic frequency rather than the
Keplerian frequency. For an infinitely narrow boundary layer, f = 2,
while for $\bar{r}/r_2$ = 0.9 and 0.8, f = 1.84 and 1.65
respectively. The resulting modified constraints are shown in
Figure~2. We conclude that a careful interpretation of the kHz QPO frequencies
makes the constraints on the masses and radii of neutron stars much less
stringent.

We note that in general relativity, the radial epicyclic frequency for
test particle orbits deviates from the Keplerian form at radii close
to $r_{\rm ISCO}$. Unlike the Newtonian case, in general relativity,
radial and vertical epicyclic frequencies do not coincide with the
azimuthal frequency.  Simple derivations of epicyclic oscillation
frequencies in the (Newtonian, as well as) general relativistic cases
are provided by Abramowicz \& Kluzniak (2004), who give expressions
for the dynamical frequencies in terms of the metric. General
relativistic expressions for the radial epicyclic frequency
incorporating fluid effects and appropriate formulations of the metric
around the neutron star will be the subject of future work. By
continuity from the Newtonian treatment here, we anticipate that the
highest frequency will turn out to be the radial epicyclic frequency
at radii $r_1, \bar{r}>r_{\rm ISCO}$ by factors of a few.

\acknowledgements

M.\,A.\,A. thanks the Physics Department of the University of Arizona, 
D.\,P.\, thanks Sabanc{\i} University, and both authors thank the University
of Amsterdam for hospitality during the writing of this paper. M.\,A.\,A.\
acknowledges support from the Turkish Academy of Sciences and the Sabanc{\i}
University Astrophysics and Space Forum. D.\,P.\ acknowledges support from the
NASA grant NAG-513374.

\clearpage 
\end{document}